\documentclass[aps,pra,twocolumn,superscriptaddress,10pt]{revtex4}
\usepackage[dvips]{graphics}
\usepackage{graphicx}
\usepackage{amsfonts}
\usepackage{amssymb}
\usepackage{amsmath}
\usepackage{color}

\newcommand{\figwidth}{7.5cm}
\newcommand{\figwidthtwo}{8.5cm}

\begin{document}

\title{Nonlinear Excitations, Stability Inversions and
Dissipative Dynamics in Quasi-one-dimensional Polariton Condensates}

\author{J.\ Cuevas}
\affiliation{Grupo de F\'{\i}sica No Lineal.  Departamento de F\'{\i}sica Aplicada I.
Escuela Polit\'ecnica Superior, Universidad de Sevilla, C/ Virgen de \'Africa, 7, 41011-Sevilla, Spain}

\author{A.S.\ Rodrigues}
\affiliation{Departamento de F\'{\i}sica/CFP, Faculdade de Ci\^{e}ncias, Universidade do Porto, R. Campo Alegre,
687 - 4169-007 Porto, Portugal}

\author{R.\ Carretero-Gonz\'alez}
\affiliation{Nonlinear Dynamical Systems Group,%
\footnote{\texttt{URL}: http://nlds.sdsu.edu}%
Department of Mathematics and Statistics, and Computational Science
Research Center, San Diego State University, San Diego CA, 92182-7720, USA}

\author{P.G.\ Kevrekidis}
\affiliation{Department of Mathematics and Statistics, University of Massachusetts,
Amherst MA 01003-4515, USA}

\author{D.J.\ Frantzeskakis}
\affiliation{Department of Physics, University of Athens, Panepistimiopolis,
Zografos, Athens 157 84, Greece}

\begin{abstract}
We consider the existence, stability and dynamics
of the ground state and nonlinear excitations, in the form of
dark solitons, for a quasi-one-dimensional polariton condensate
in the presence of pumping and nonlinear damping.
We find a series of remarkable features that can be directly contrasted
to the case of the typically energy-conserving ultracold alkali-atom
Bose-Einstein condensates. For some sizeable parameter ranges,
the nodeless (``ground'') state becomes {\it unstable}
towards the formation of {\em stable} nonlinear single or {\em multi}
dark-soliton excitations. It is also observed that for suitable
parametric choices, the instability of single dark solitons
can nucleate multi-dark-soliton states.
Also, for other parametric regions, {\em stable asymmetric} sawtooth-like
solutions exist. Finally, we
consider the dragging of a defect through the condensate and the 
interference of two initially separated condensates, both of which
are capable of nucleating dark multi-soliton dynamical states.
\end{abstract}
\date{\today}

\maketitle

\section{Introduction}

An important recent development that has spurted a new direction for
the physics of Bose-Einstein condensation
has been the observation of condensation phenomena for
polaritons in semiconductor microcavities \cite{kasp1_and_more} at much higher
temperatures than ultracold atomic Bose-Einstein condensates (BECs) \cite{stringari1,stringari2}.
In the setting of exciton-polariton BECs, the condensing ``entities'' are excitons,
i.e., bound electron-hole particles. When confined, these
develop strong coupling
with light, forming exciton-photon mixed quasi-particles known
as polaritons \cite{micro_cavity_polaritons}.
The finite temperature formation of polariton condensates
leads the quasi-particles to also possess a finite lifetime:
in fact, they can only exist  for a few picoseconds in the cavity
before they decay into photons. Hence, thermal equilibrium
can never be achieved and the system produces a genuinely
far-from-equilibrium condensate, in which external pumping from a reservoir
of excitons counter balances the loss of polaritons due to the above decay
mechanism. Nevertheless,
numerous key features of the superfluid character of the exciton-polariton condensates
have been established, including the flow without scattering (analog of the
flow without friction) \cite{amo1}, the existence of vortices \cite{lagou1}
(see also Ref.~\cite{roumpos} for vortex dipole dynamics),
the collective dynamics \cite{amo2}, as well as remarkable applications
such as spin switches \cite{amo3} and light emitting diodes \cite{amo4}
operating even near room temperatures.

The pumping and damping mechanisms associated with polaritons enable
the formulation of different types of models. One of these, proposed
in Refs.~\cite{berloff1,kbb} suggests the use of a single partial
differential equation (PDE) for the polariton
condensate incorporating the above mentioned loss-gain mechanisms.
This model features a localized (within a
pumping region) gain and a nonlinear saturating loss of
polaritons; these are the fundamental differences
of this setting from the standard PDE mean-field model, namely the
Gross-Pitaevskii equation (GPE) used in the physics of atomic
BECs \cite{stringari1,stringari2,emergent}.
In another class of models, which has been
proposed in Refs.~\cite{polar1,polar2,cc05},
the polaritons are coupled to the evolution of the exciton population;
such models
also display nonlinear diffusive spatial dynamics for the excitons.

In this work, our aim
is to consider the quasi-one-dimensional (1D) dynamics of
polariton BECs and to illustrate their {\it fundamental
differences} from the more standard alkali-atom condensates.
At this point we should mention that polariton condensates considered
so far (also experimentally) have been intrinsically
two-dimensional (2D) \cite{RMP}. Nevertheless,
using tight confinement along one, transversal,
direction
(i.e., highly anisotropic variants of the traditional parabolic traps)
we envision rendering the polariton condensate effectively
1D \cite{1d_polaritons}.
In addition, a broader perspective for our considerations
is that understanding the nonlinear dynamics
and pertinent phenomenology in the 1D setting, may pave the way
towards subsequently generalizing relevant considerations to the more realistic 2D case.
The key phenomena that are reported herein are: a wide parametric interval
of destabilization of the fundamental nodeless state
of polariton BECs;
a partial stability within this interval of excited states, in the form of dark solitons;
the spontaneous production of higher excited (multi-soliton)
states from lower ones or even from nodeless states; the production
of dark soliton trains by dragging of a defect through the polariton
BEC (cf.~the recent relevant 2D experimental results in Ref.~\cite{aamo});
and finally the formation of long-lived dark multi-soliton dynamical
states through the interference of two separated polaritonic 
condensates in analogy with the atomic BEC case of Ref.~\cite{weller}.

The paper is organized as follows. In Section II we describe our model and setup,
providing also a brief description of our methods. Section III is devoted to
our detailed numerical investigations, and Section IV concludes our work,
including some suggestions for possible future studies.

\section{Model setup}

In our analysis below, we consider
the modified complex Gross-Pitaevskii model developed in Refs.~\cite{berloff1,kbb}
suitably reduced to one spatial dimension:
\begin{equation}
\label{eq:dyn}
i\partial_t\psi=
\left\{-\partial_{x}^2+x^2+|\psi|^2+i\left[(\chi(x)-\sigma|\psi|^2)\right]\right\}\psi,
\end{equation}
where $\psi$ denotes the polariton wavefunction trapped inside
a 1D
harmonic potential, $x^2$ (the transverse direction,
perpendicular to $x$, corresponds to the tight trapping
axis mentioned above).
The differences of Eq.~(\ref{eq:dyn}) from the
standard GPE appearing in the physics of atomic BECs can be traced
to the presence of (i) the spatially dependent gain term with
\begin{equation}
\chi(x)=\alpha\Theta(x_m-|x|),
\end{equation}
where $\Theta$ is the step function generating a symmetric spot
of ``radius'' $x_m$ and strength $\alpha$ for the gain
and (ii) the nonlinear saturation loss term
of strength $\sigma$.
Estimates of the relevant physical time and space scales, as well as physically relevant
parameter values, are given in Ref.~\cite{berloff1}. We should
also note in passing that although our results below are given in the
context of Eq.~(\ref{eq:dyn}), we have ensured that similar phenomenology
arises in the model of Refs.~\cite{polar1,polar2,cc05}, for suitable
parametric choices.
In other words, the phenomenology that is reported in this work
is {\it generically} relevant to (1D) polariton BECs independently of model
specifics.

In what follows, we will consider the
stationary solutions of
the quasi-1D model at hand, in the form
$\psi(x,t) = \psi_0(x)\exp(-i \mu t)$
where $\mu$ is the dimensionless chemical potential, and
the stationary state $\psi_0(x)$ is governed by
the following ordinary differential equation:
\begin{equation}
\label{eq:stat}
\mu\psi_0=\left\{-\frac{d^2}{dx^2}
    +x^2+|\psi_0|^2+i\left[(\chi(x)-\sigma|\psi_0|^2)\right]\right\}\psi_0.
\end{equation}
%
Importantly, the additional
condition
\begin{equation}
    \int \mathrm{d}x\,(\chi(x)-\sigma|\psi|^2)|\psi_0|^2=0
\label{eq:stat1}
\end{equation}
needs to be enforced as a population balance constraint.
This self-consistently selects the particular value of the chemical
potential once the other parameters (i.e., $\alpha$, $\sigma$, $x_m$)
are fixed. This is why some of our graphs of the solution branches
below will feature $\mu$ as a function of other solution parameters,
such as $x_m$.
We note in passing the significant difference of this trait from
the Hamiltonian atomic BEC case, where there exist monoparmetric
families of solutions as a function of $\mu$. 
Once stationary solutions of the differential-algebraic
system of Eqs.~(\ref{eq:stat})-(\ref{eq:stat1}) are identified,
their linear stability is considered by means of a Bogolyubov-de Gennes
analysis. Namely, small perturbations 
[of order ${\rm O}(\delta)$, with $0< \delta \ll 1$]
are introduced in the form
$$
\psi(x,t)=e^{-i \mu t} \left[\psi_0(x) + \delta
(a(x) e^{i \omega t} + b^{*}(x) e^{-i \omega^{*} t}) \right],
$$
and the ensuing linearized equation
are then solved to O$(\delta)$,
leading to the following eigenvalue problem:
%
\begin{equation}
    \omega
    \left(\begin{array}{c} a(x) \\ b(x) \end{array}\right)=
    \left(\begin{array}{cc} L_1 & L_2 \\ -L_2^* & -L_1^* \end{array}\right)
    \left(\begin{array}{c} a(x) \\ b(x) \end{array}\right),
\end{equation}
for the eigenvalue $\omega$
and associated eigenvector $(a(x),b(x))^T$, where
$L_1$ and $L_2$ are the following operators:
\begin{eqnarray}
    L_1&\!=\!&-\mu-\frac{d^2}{dx^2}
    +x^2+2(1-i\sigma)|\psi_0|^2+i\chi(x),
\notag
\\[1.0ex]
\notag
    L_2&\!=\!&(1-i\sigma)\psi_0^2.
\end{eqnarray}

\begin{figure}
\begin{center}
    \includegraphics[width=\figwidth]{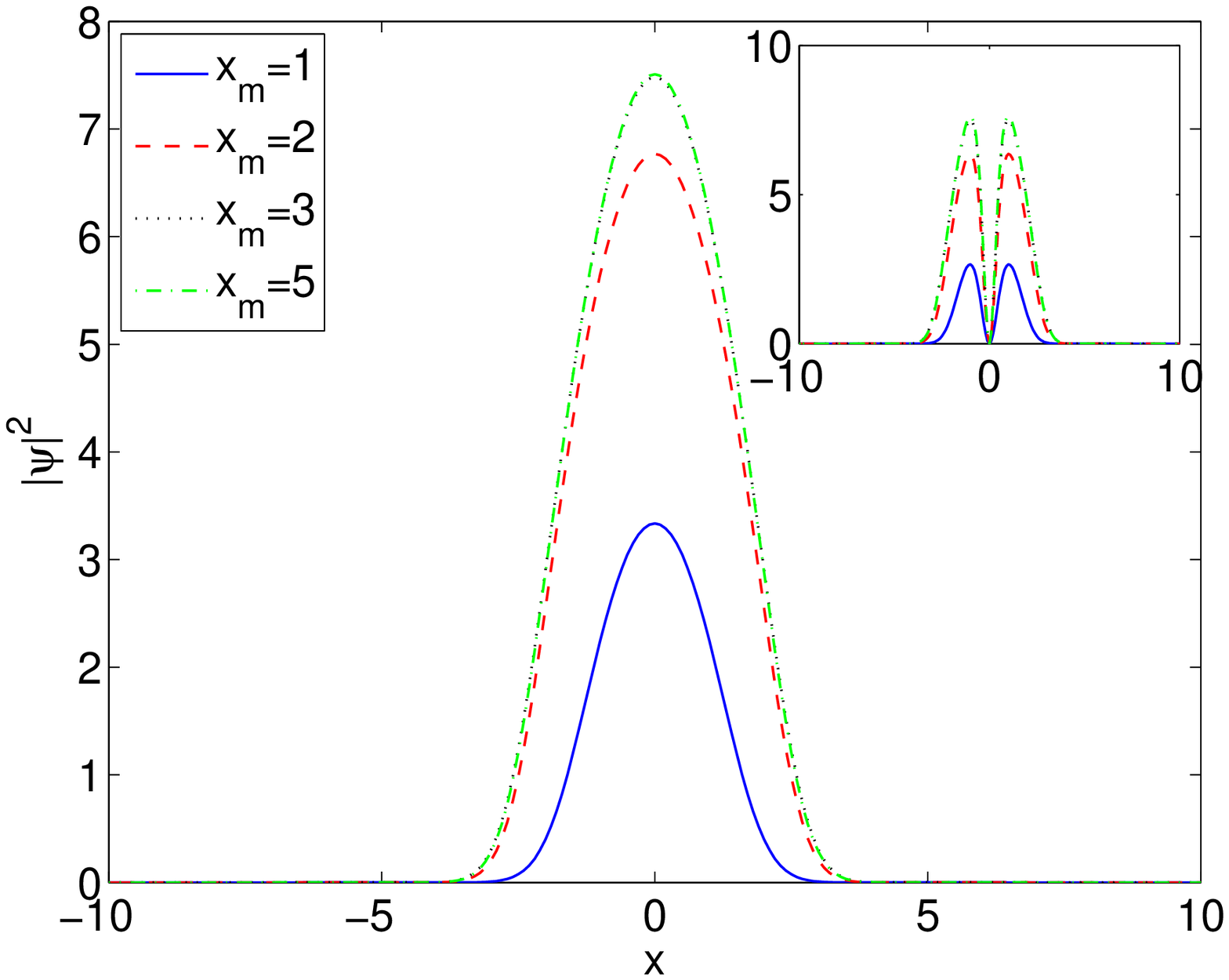}
    \includegraphics[width=\figwidth]{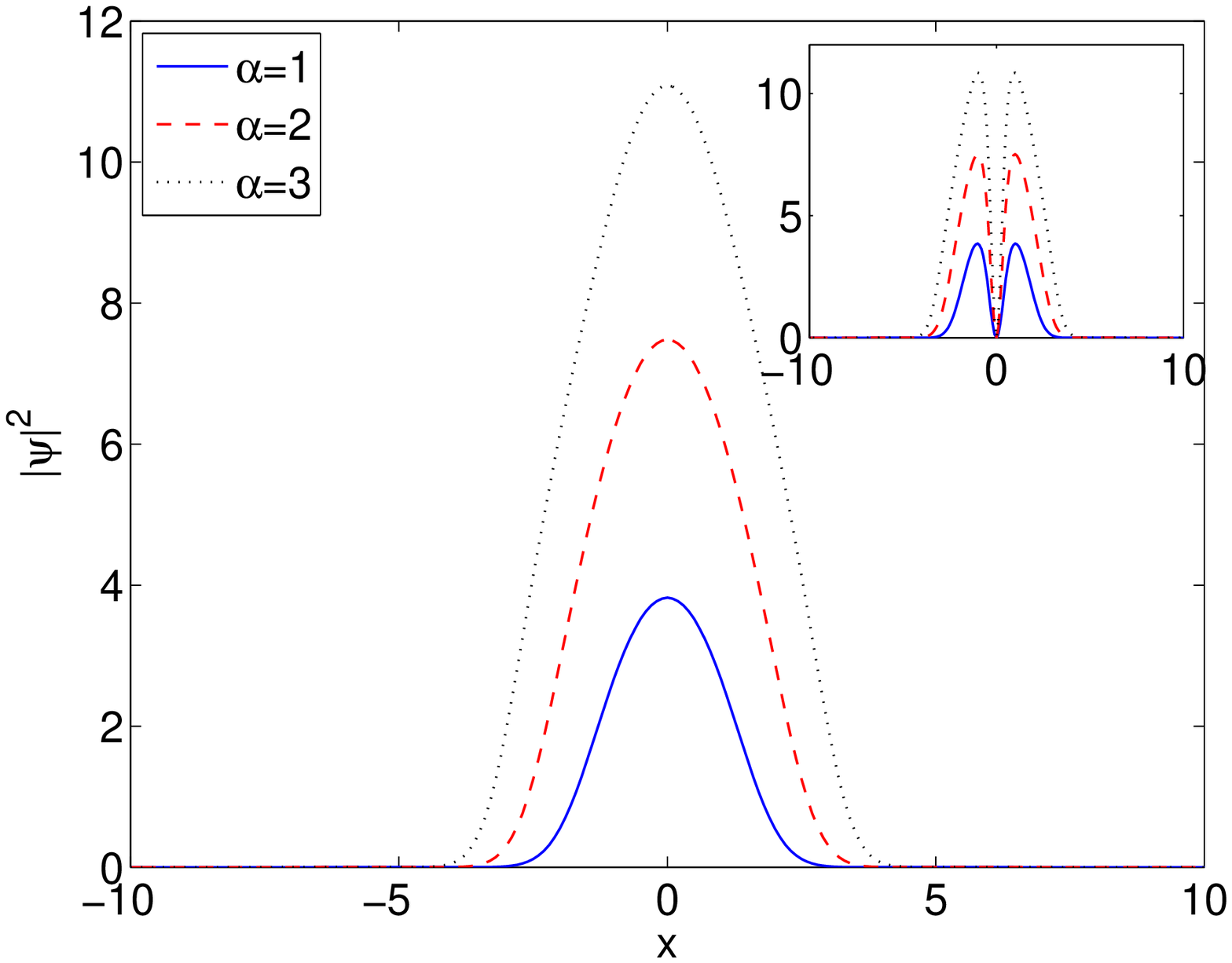}
\caption{(Color online) Spatial profiles of the densities, $|\psi|^2$,
of nodeless states
and single dark solitons (insets).
In the top (bottom) panel, the parameter $\alpha$ ($x_m$) is kept fixed,
taking the value $\alpha=2$ ($x_m=3$), while
$\sigma=3.5$ in both cases.
As seen in the top panel, the profiles do not change appreciably above
a critical value of $x_m\approx3$: in fact, the profiles corresponding to
$x_m=3$ [dotted (black) line]
and $x_m=5$ [dotted-dashed (green) line]
are almost indistinguishable from each other.
All quantities in this and in all subsequent figures are dimensionless.}
\label{fig:profiles1}
\end{center}
\end{figure}

Once the stationary solutions are found to be linearly unstable (i.e., ${\rm Im}\{\omega\} \ne 0$),
then the dynamical manifestation of the corresponding instabilities
is monitored through direct numerical simulations of Eq.~(\ref{eq:dyn}).
%
%
%
%
%
%
%
%
%
In addition, in what follows, we have considered dynamical scenarios
under which nonlinear excitations, such as single- or multiple-dark-solitons
\cite{emergent} can arise in the context
of polariton BECs. Such excited states have been extensively studied
in the context of atomic BECs \cite{djf}, while they
have been amply considered in recent experimental investigations in this context
\cite{weller,sengstock,jeff}.
Additionally, motivated by relevant studies in atomic BECs \cite{hau,engels} and recent
experiments in polariton condensates \cite{amo1,aamo,cancellieri}, we consider
the nucleation of dark solitons
by a moving defect, modeled by a (localized) potential; the latter, is assumed to be
produced by a narrow laser beam of Gaussian shape, namely:
\begin{equation}
    V_{\rm def}=V_0\exp(-(x-vt)^2/\epsilon^2),
    \label{obs}
\end{equation}
where $V_0$, $v$ and $\epsilon$ represent the amplitude, speed, and width of
the potential, respectively. We consider the fixed point solution of
the modified GPE (\ref{eq:dyn}) in the presence of this second defect potential
(in addition to the harmonic trap $x^2$)
at the center of the trap. Then we evolve the system in time
starting from this solution, dragging the defect through the system.
This is similar in spirit to the recent experiments of Ref.~\cite{amo1,aamo} (and to
the recent theoretical investigation of Ref.~\cite{wou10}).
Finally, we study an alternative proposal for dark soliton nucleation,
akin to the interference experiments conducted for atomic
BECs in Ref.~\cite{weller}, whereby a central potential barrier
separating two polariton clouds is lifted
allowing the two clouds to interfere; this process leads
to the production of persistent dark solitons as well.

\begin{figure}
\begin{center}
    \includegraphics[width=\figwidth]{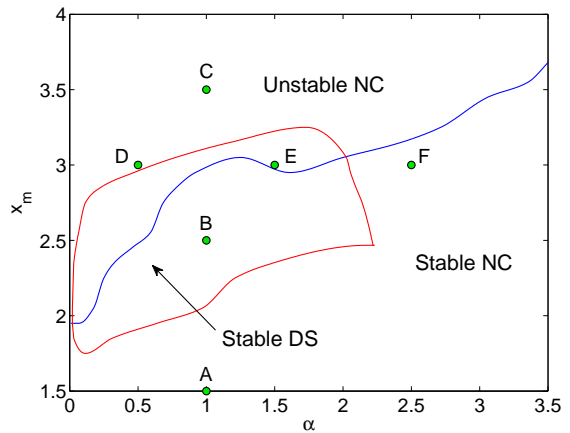}
\caption{(Color online) Stability domains
of nodeless clouds (NC) and dark solitons (DS) for $\sigma=0.35$ and $\alpha\leq3.5$. Dark solitons are stable only in the area indicated by the arrow. Notice that there is a superposition between part of the dark soliton and nodeless cloud stability ranges.
The (green) circles indicate the parameter locations for the spectra
depicted in Fig.~\ref{fig:stab}.
}
\label{fig:ranges1}
\end{center}
\end{figure}

\begin{figure}
\begin{center}
    \includegraphics[width=\figwidthtwo]{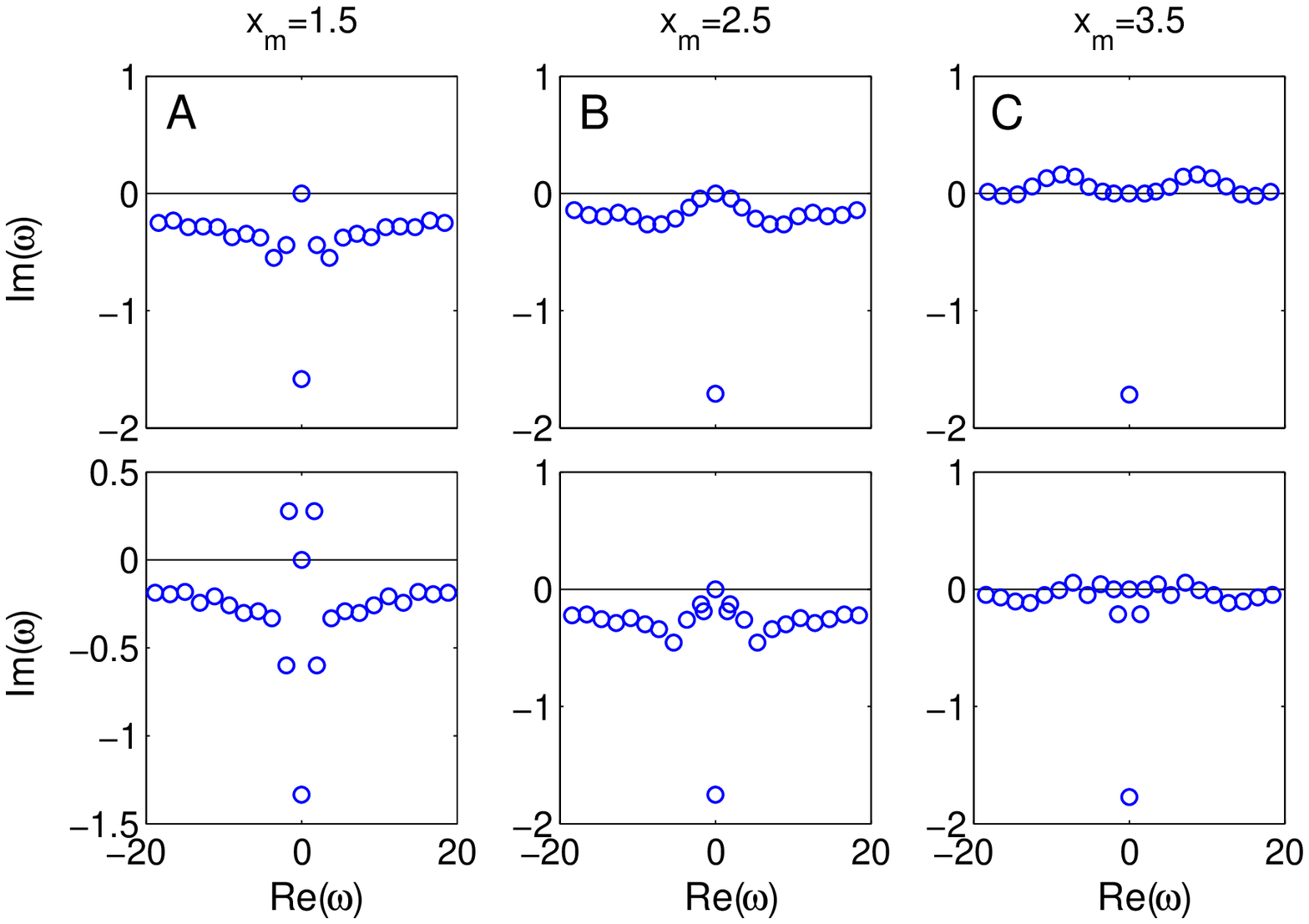}
\\[3.0ex]
    \includegraphics[width=\figwidthtwo]{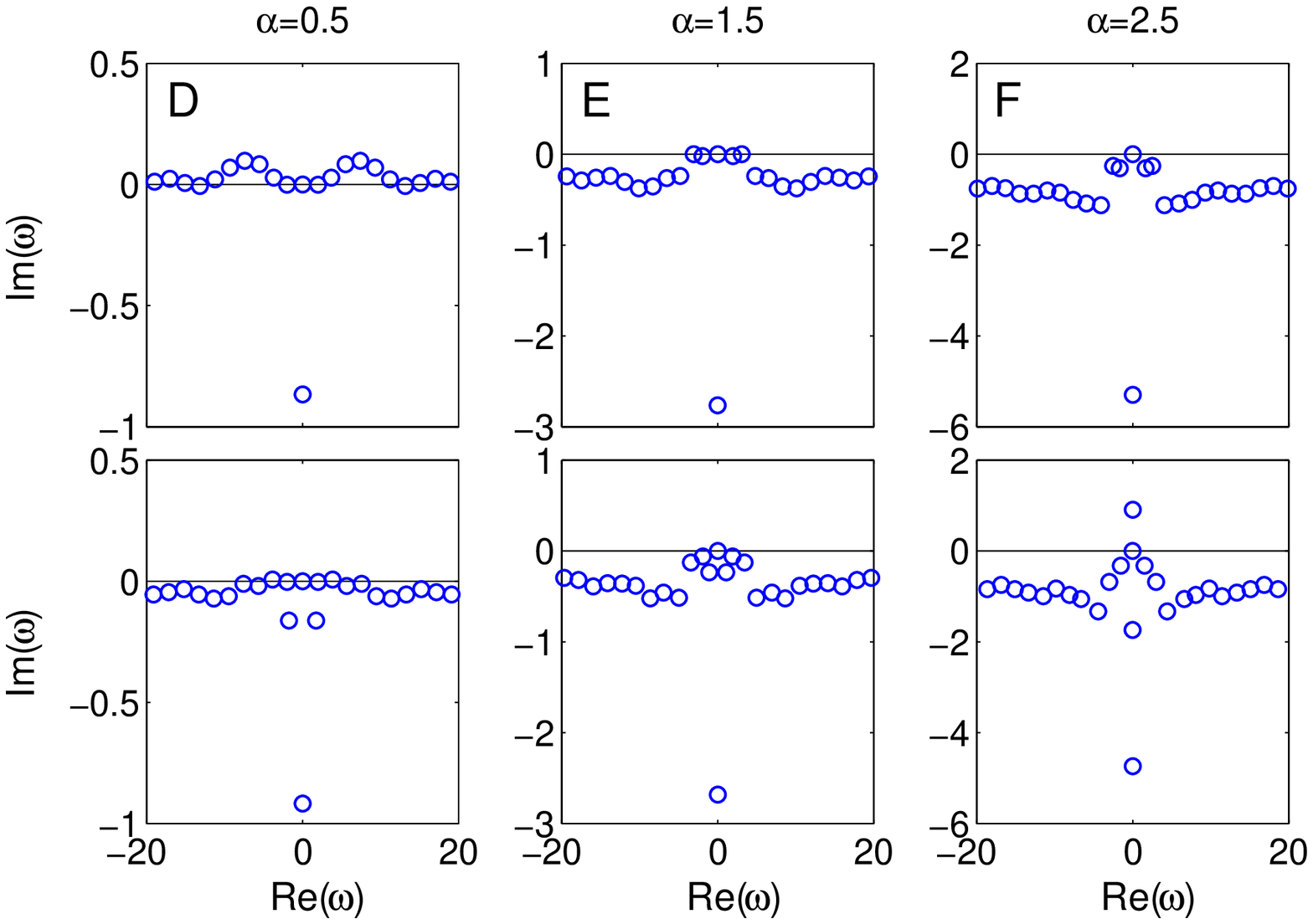}
\caption{(Color online) Spectral plane for nodeless clouds (top row in each series of panels)
and dark solitons (bottom row in each series of panels) with $\sigma=0.35$.
%
%
The top series of panels correspond to fixed $\alpha=1$ and
increasing values of $x_m$ as labeled
whereas the bottom series of panels correspond to fixed $x_m=3$ and
increasing values of $\alpha$ as indicated.
The different cases correspond to the parameter locations depicted
by the (green) circles in Fig.~\ref{fig:ranges1}.
}
\label{fig:stab}
\end{center}
\end{figure}

\begin{figure}
\begin{center}
    \includegraphics[width=\figwidth]{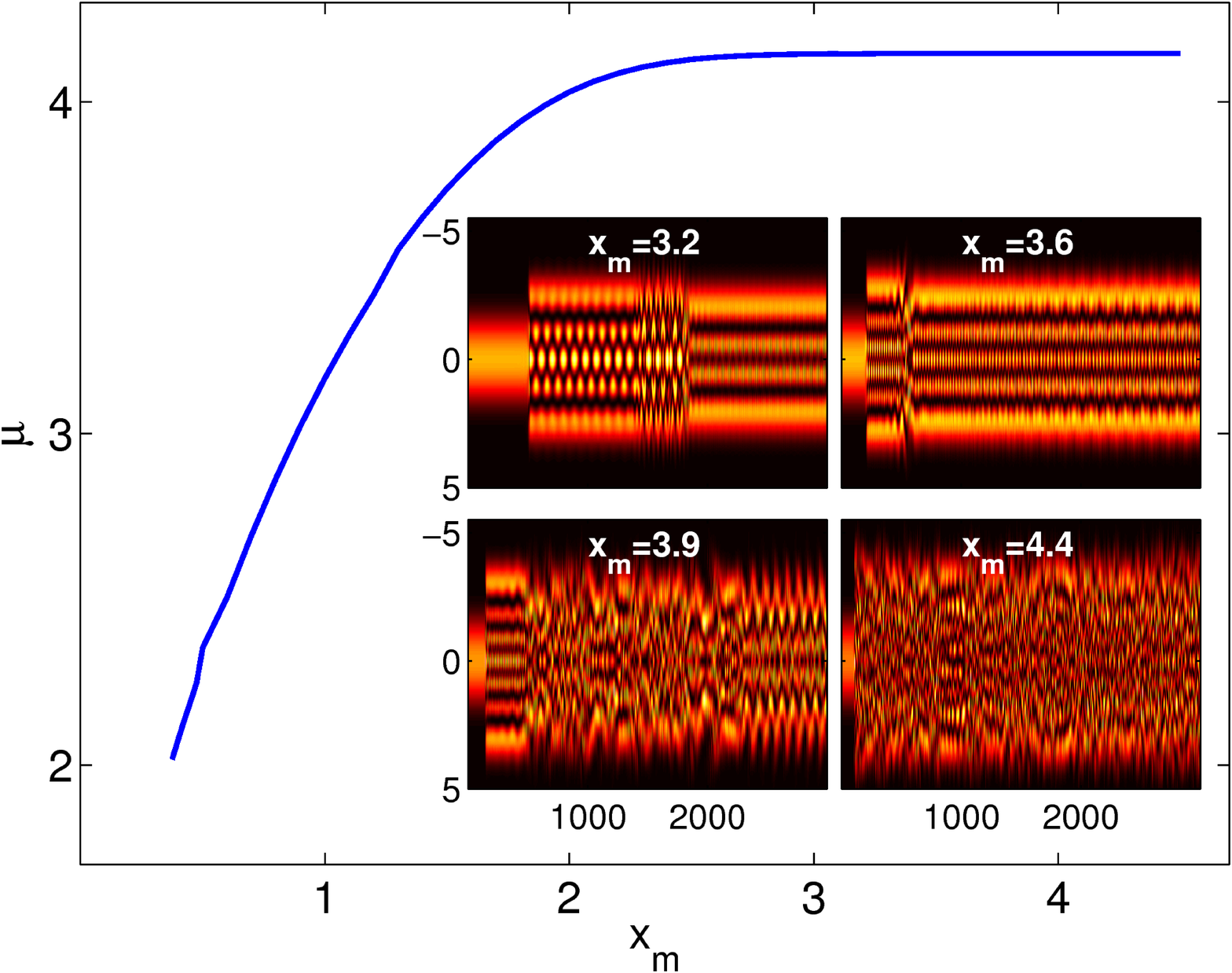}
\caption{(Color online) Top panel: Chemical potential as a function of $x_m$ for a
nodeless cloud with $\alpha=1.0$ and $\sigma=0.35$. The insets
show the time evolution of the cloud for various values of $x_m$.
}
\label{fig:dyn1}
\end{center}
\end{figure}

\begin{figure}
\begin{center}
    \includegraphics[width=\figwidth]{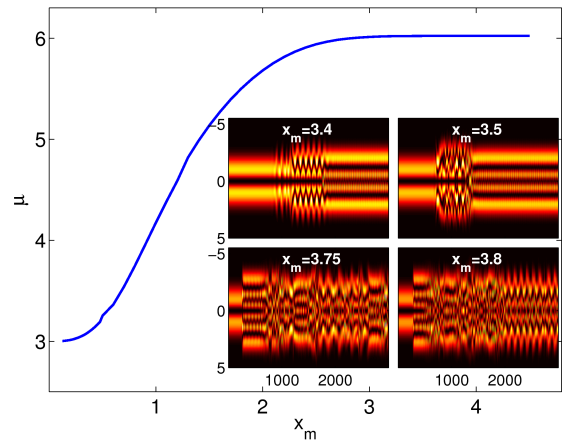}
\caption{(Color online) Top panel:
Chemical potential as a function of $x_m$ for a dark soliton with $\alpha=1.0$ and $\sigma=0.35$.
The insets show the time evolution of the cloud for various values of $x_m$.
}
\label{fig:dyn2}
\end{center}
\end{figure}

\section{Numerical results}
We hereby explore the existence, stability and dynamical
properties of the nodeless cloud (NC) as well as of excited states exhibiting a single-node,
namely dark solitons (DSs). The above mentioned states (NC and DS) are
the most fundamental nonlinear states of the system,
whose profiles ---for different parameter combinations--- are displayed in
Fig.~\ref{fig:profiles1}.
%
In particular, the top panel of Fig.~\ref{fig:profiles1} depicts the
NC and DS profiles for a constant saturation coefficient $\sigma=3.5$ and
constant pumping spot strength $\alpha=2$, and a
varying
radius of the pumping spot $x_m$.
We observe that, for given values of $\alpha$ and $\sigma$,
there is a critical value of $x_m$ above which the shapes of
the NCs and DSs remain unchanged.
In the example depicted in
the top panel of Fig.~\ref{fig:profiles1}, the profiles for $x_m \gtrsim 3$
are indistinguishable from each other (see profiles for $x_m=3$ and $x_m=5$).
This effect is due to the fact that for large $x_m$, the pumping spot
covers a larger portion than the saturated size of the cloud
when loss and gain are balanced. This cloud size is the analogue
of the Thomas-Fermi radius for an atomic condensate.
The saturation of the cloud size is equivalent to the
saturation of the chemical potential $\mu$ as $x_m$ is increased,
as
shown below (see Figs.~\ref{fig:dyn1} and \ref{fig:dyn2}).
In the bottom panel of Fig.~\ref{fig:profiles1} we depict the NC
and DS profiles for a constant pumping spot radius $x_m=3$ and a varying
pumping spot strength. Note that, in this case, the size of the
cloud continuously expands (in amplitude and width) with increasing 
spot strength.

We now proceed to provide a characterization of the existence
and stability properties of the NC and DS profiles with respect
to the various parameters at hand.
In what follows, in order to offer
a picture of the relevant parameter space, we have
varied the gain parameters $\alpha$ and $x_m$ whereas the coefficient of
the saturating nonlinear loss $\sigma=0.35$ has been kept fixed.
Figure~\ref{fig:ranges1} depicts the existence and stability domains
in the $(\alpha,x_m)$ parameter
plane with
$\sigma=0.35$ fixed for both the nodeless cloud
and the single DS that can be found as (numerically exact up to a
prescribed accuracy) fixed point solutions of Eq.~(\ref{eq:stat}).
%
The NC and DS configurations exist for all parameter combinations,
as it is the case for atomic condensates.
%

Nevertheless, as far as the stability
and dynamical properties of NC and DS states are concerned,
we can observe {\it fundamental} differences
between the pumped-damped polariton BECs
and atomic BECs.
In particular, the nodeless cloud (which was
{\it always} stable in the Hamiltonian case of atomic BECs \cite{emergent})
{\it is now stable only below a critical value of the pumping spot size}
$x_m$. On the other hand, also remarkably, even the single DS is
stable only in a limited range, while it was {\it always} stable in
quasi-1d
harmonically trapped atomic BECs (see, e.g., Ref.~\cite{weller} and
references therein). Moreover, in a very unusual manifestation of stability
inversions, not only can the nodeless state be stable
while the dark soliton is not, but also vice versa: the state with
a node can be stable
while the one without a node is not.
In Fig.~\ref{fig:stab} we show the details of the Bogolyubov spectra
of both states. These showcase the dissipative nature of the dynamics
being associated with frequencies chiefly with negative imaginary
part; moreover,
they also illustrate the potential instabilities arising in the
system through Hopf bifurcations and oscillatory instabilities associated
with complex eigenfrequencies, or through zero crossings (and purely
imaginary eigenfrequencies). This second scenario only appears for
dark solitons. From an intuitive viewpoint, this phenomenology can
partially be understood on the following grounds. The condensate
in the absence of the external driving has an intrinsic length scale
selected by the trap (and the chemical potential). The presence
of the external forcing over the radius $x_m$ introduces an 
additional length scale competing with the former one. Hence, when
this forcing becomes fairly (spatially) extended, it favors a spatially
wider state. This is manifested through the instability of a group of 
``background'' spectral modes (which are not the lowest modes of the
condensate close to the spectral plane origin) in the panels of 
Fig. \ref{fig:stab}. On the contrary, the dark soliton or multi-soliton
states may become unstable through the same mechanism, 
or they may also become unstable through their ``internal
modes'' \cite{weller} which lead to the isolated instability through the
zero crossing.



In Fig.~\ref{fig:dyn1} we show a continuation 
of the nodeless state, for a fixed value of $\alpha=1.0$. In 
particular, we depict
the chemical potential $\mu$ as a function of pumping spot size, $x_m$.
The insets show the dynamical evolution of the fixed point solution for
several illustrative values of $x_m$, which reflect the different
possible dynamical phenomena. The values shown are in the unstable
domain of the NC.
%
%
There are two types of behavior.
The first behavior, for $3.1<x_m<3.8$, corresponds to the NC decaying
into multiple-dark-solitons, by means of oscillatory transients
(even in the region of stability of the single DS).
The spontaneous emergence of
these states from a nodeless one is a feature particular to
polariton BECs, having no analog in the atomic BEC case.
On the other hand, a second behavior,
corresponds to values $x_m>3.8$: in this case, even though transient multi-soliton
states still appear, they finally give rise to nearly ``turbulent''
nonlinear dynamics of a ``sea'' of multiple DS states, which may
(or may not, depending on $x_m$) settle on an asymptotic multi-soliton
state.
%

\begin{figure}
\begin{center}
    \includegraphics[width=\figwidth]{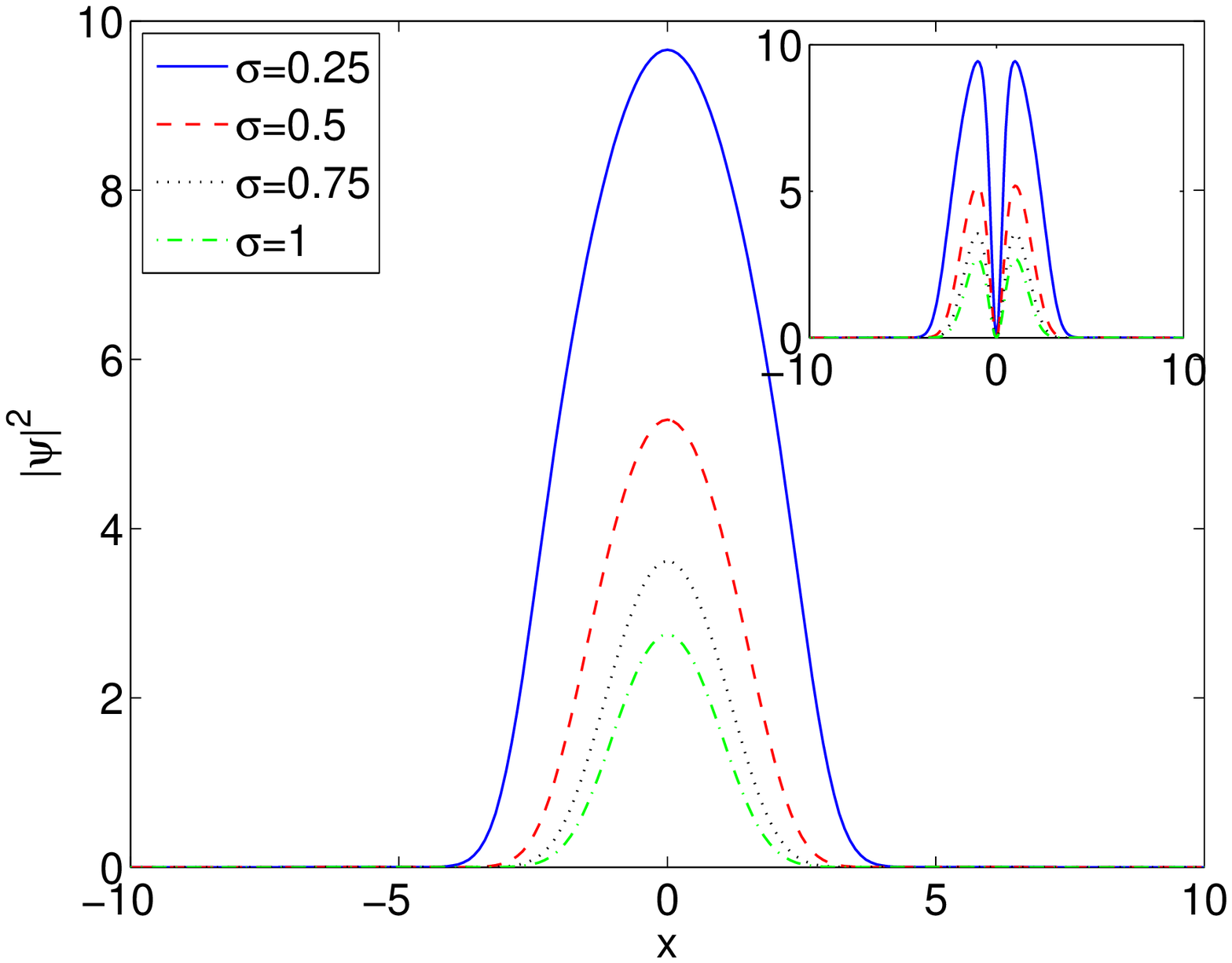}
    \includegraphics[width=\figwidth]{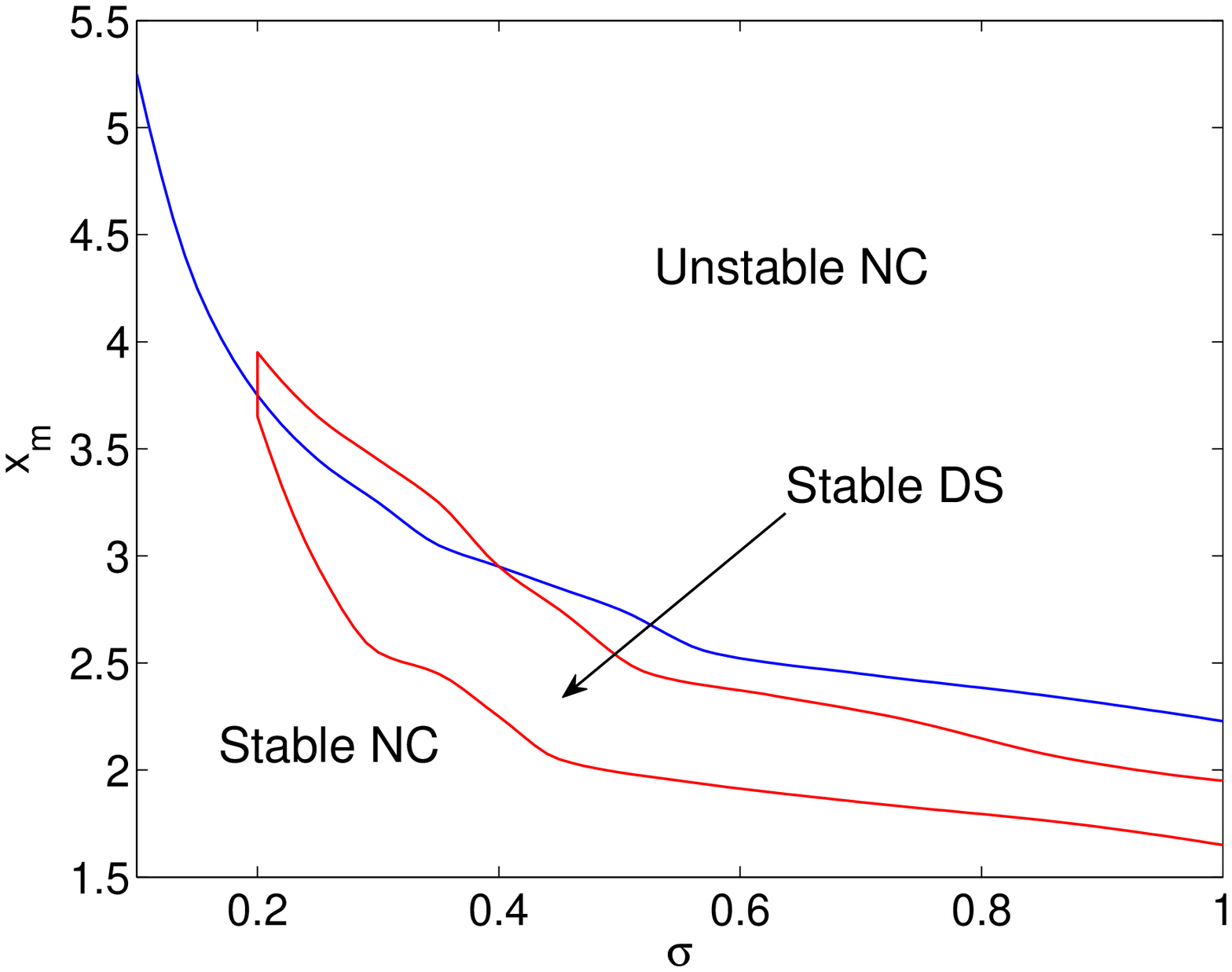}
\caption{(Color online) Top panel: Profiles for the nodeless clouds and dark solitons (inset) for different values of $\sigma$ and fixed $\alpha=2$ and $x_m=2.5$.
Bottom panel: Stability
domain for nodeless clouds (NC) and dark solitons (DS) for $\alpha=2$.
}
\label{fig:sigma1}
\end{center}
\end{figure}

\begin{figure}
\begin{center}
    \includegraphics[width=\figwidth]{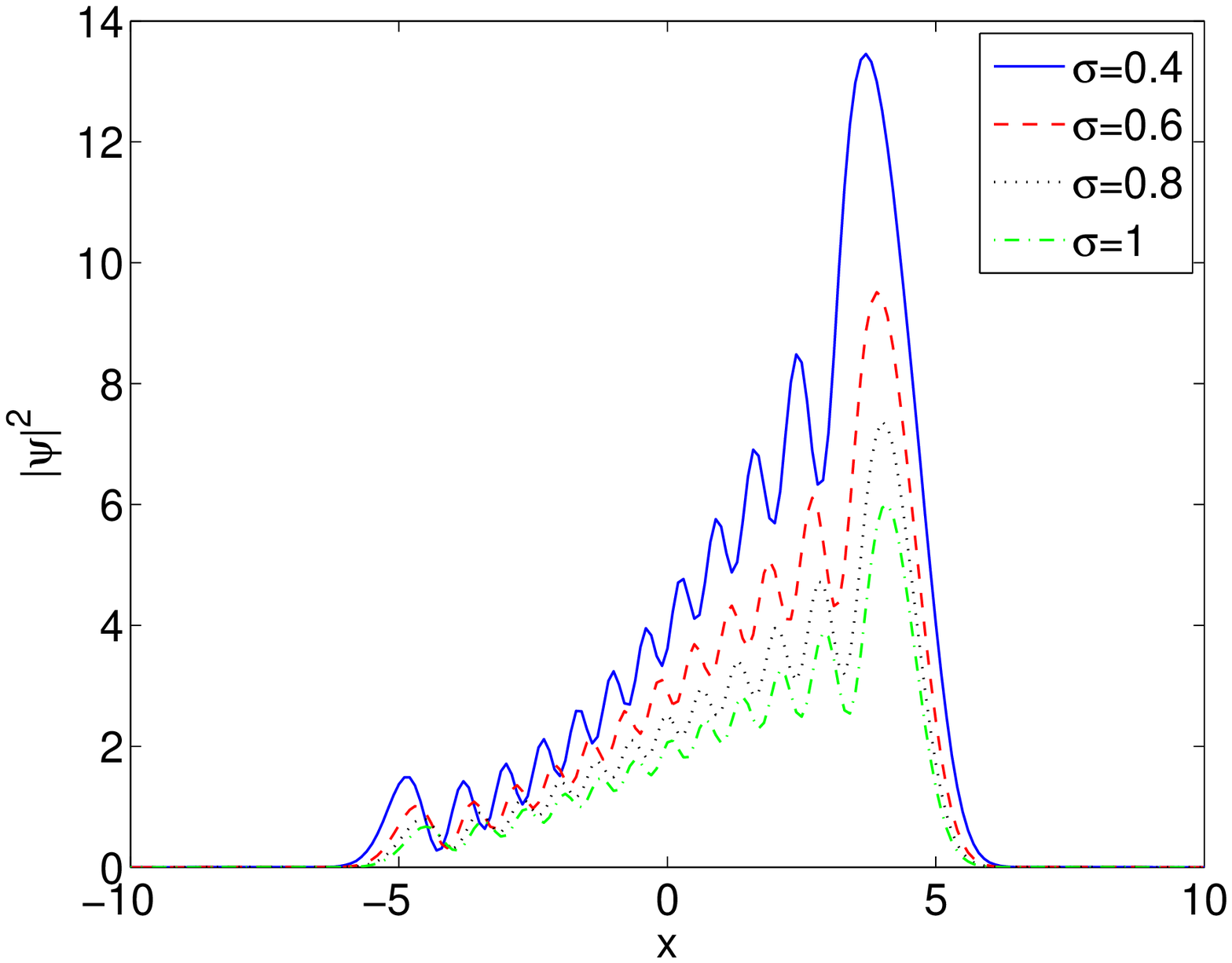}
    \includegraphics[width=\figwidth]{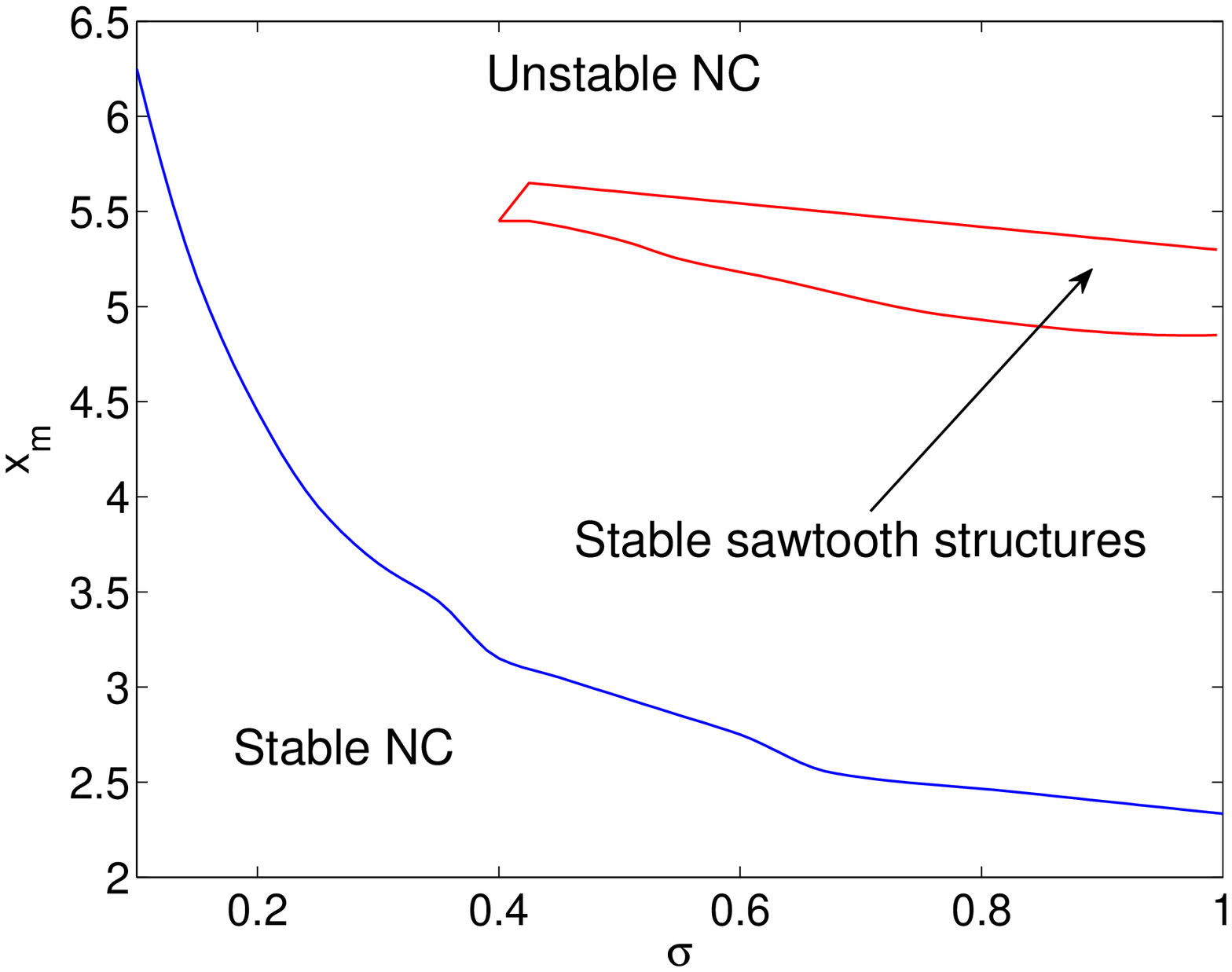}
\caption{(Color online) Top panel:
Profiles of highly asymmetric ``sawtooth'' structures
for different values of $\sigma$ and fixed $\alpha=3$ and $x_m=5$.
Bottom panel: Stability
domain for
sawtooth structures with $\alpha=3$.
}
\label{fig:sigma2}
\end{center}
\end{figure}


\begin{figure}
\begin{center}
    \includegraphics[width=\figwidth]{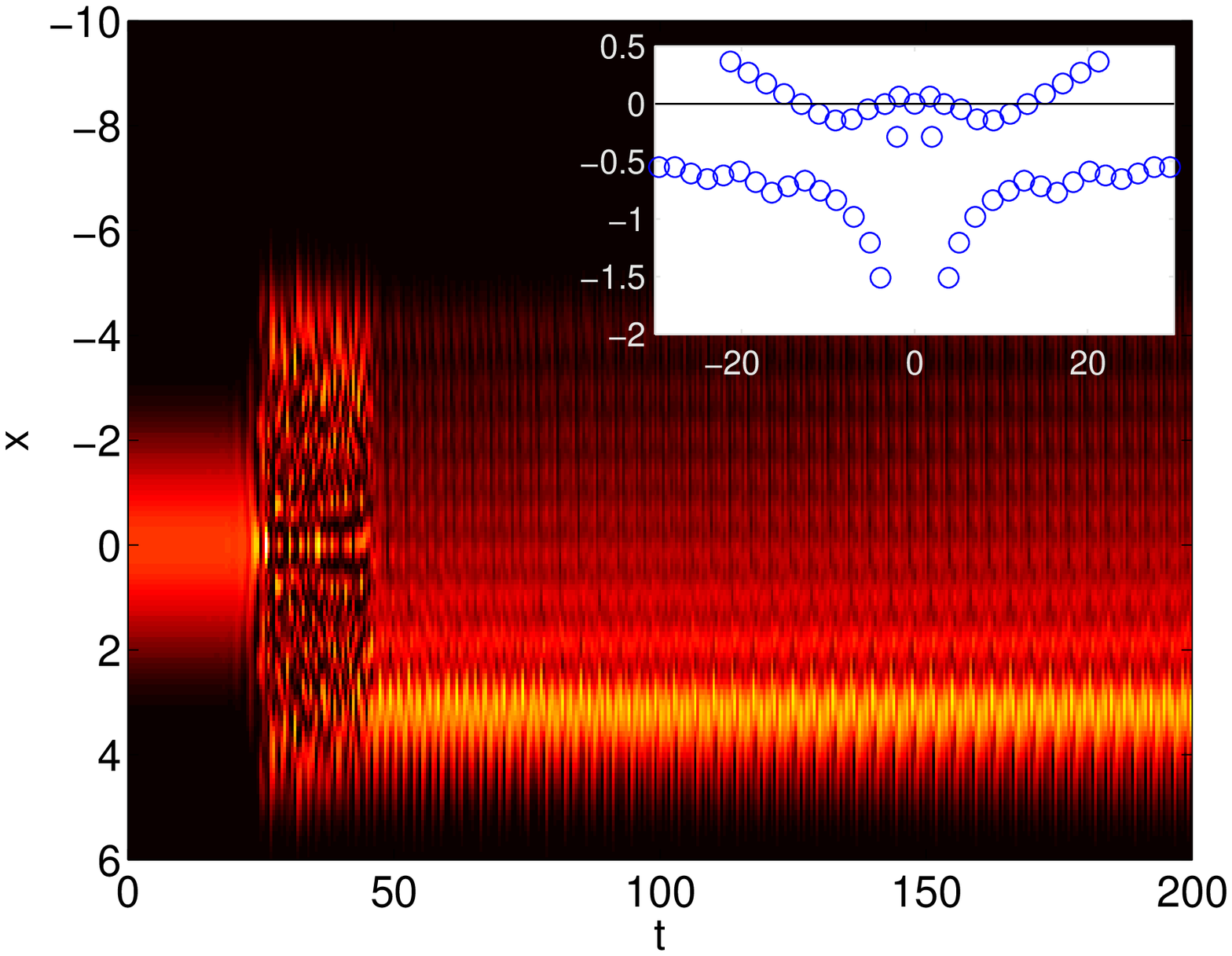}
    \includegraphics[width=\figwidth]{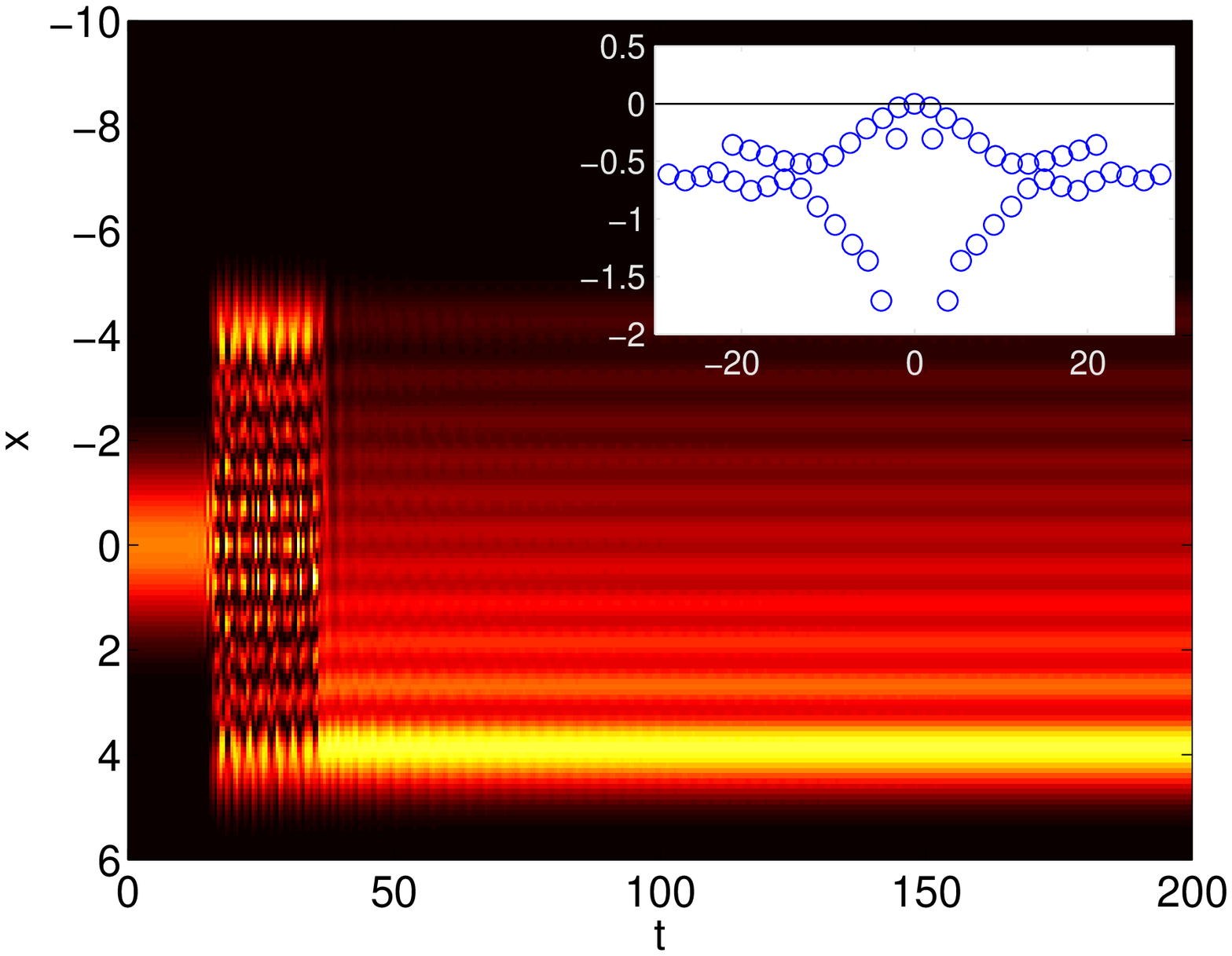}
\caption{(Color online) 
Density plots showing the evolution of unstable nodeless clouds to
a breathing sawtooth structure for $\sigma=0.5$ (top) and
a non-breathing sawtooth structure for $\sigma=1$ (bottom) with $x_m=5$.
The inset in the top panel shows the linearization spectrum of the {\em unstable} 
sawtooth steady state about which the system oscillates about
for long times.
The inset in the bottom panel corresponds to the linearization spectrum for the
{\em stable} sawtooth steady state that the system asymptotes to.
}
\label{fig:dynfinger}
\end{center}
\end{figure}

On the other hand, we have also investigated the dynamics of
the fundamental (single) dark soliton in Fig.~\ref{fig:dyn2}.
We have found
that, in their instability region, DSs decay towards
the nodeless state, as expected, if the latter is stable.
However, when the nodeless
state is unstable, the dynamics is as follows:
after a transient stage,
a breathing multiple-DS structure is formed, which may consist of
3, 4 or 5 DSs.
Examples of such evolutions
are shown in the insets of Fig.~\ref{fig:dyn2}.
In this case, the DS is stable in the range $2.1<x_m<3.1$.
%
%

\begin{figure}
\begin{center}
    \includegraphics[width=\figwidth]{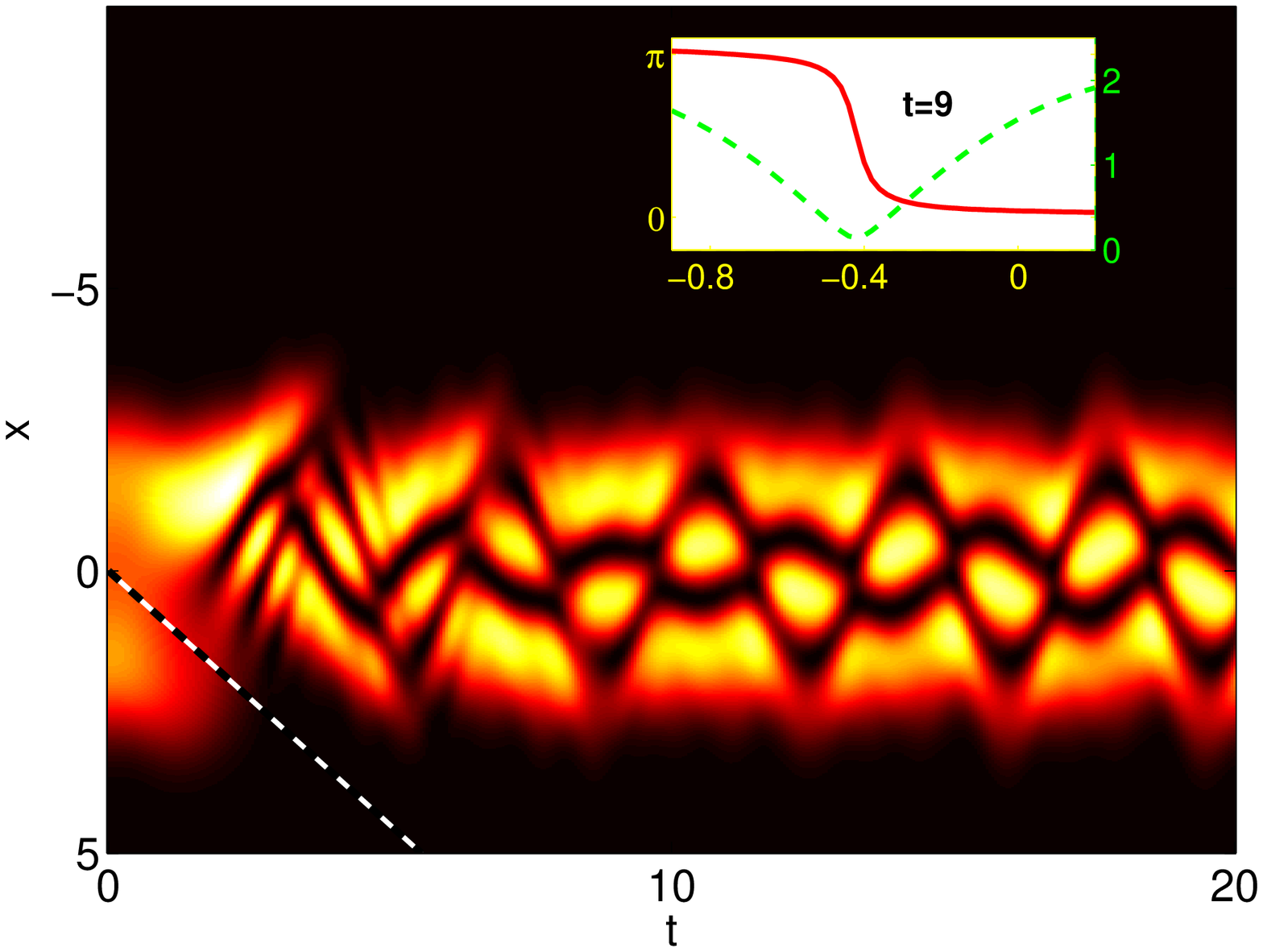}
    \includegraphics[width=\figwidth]{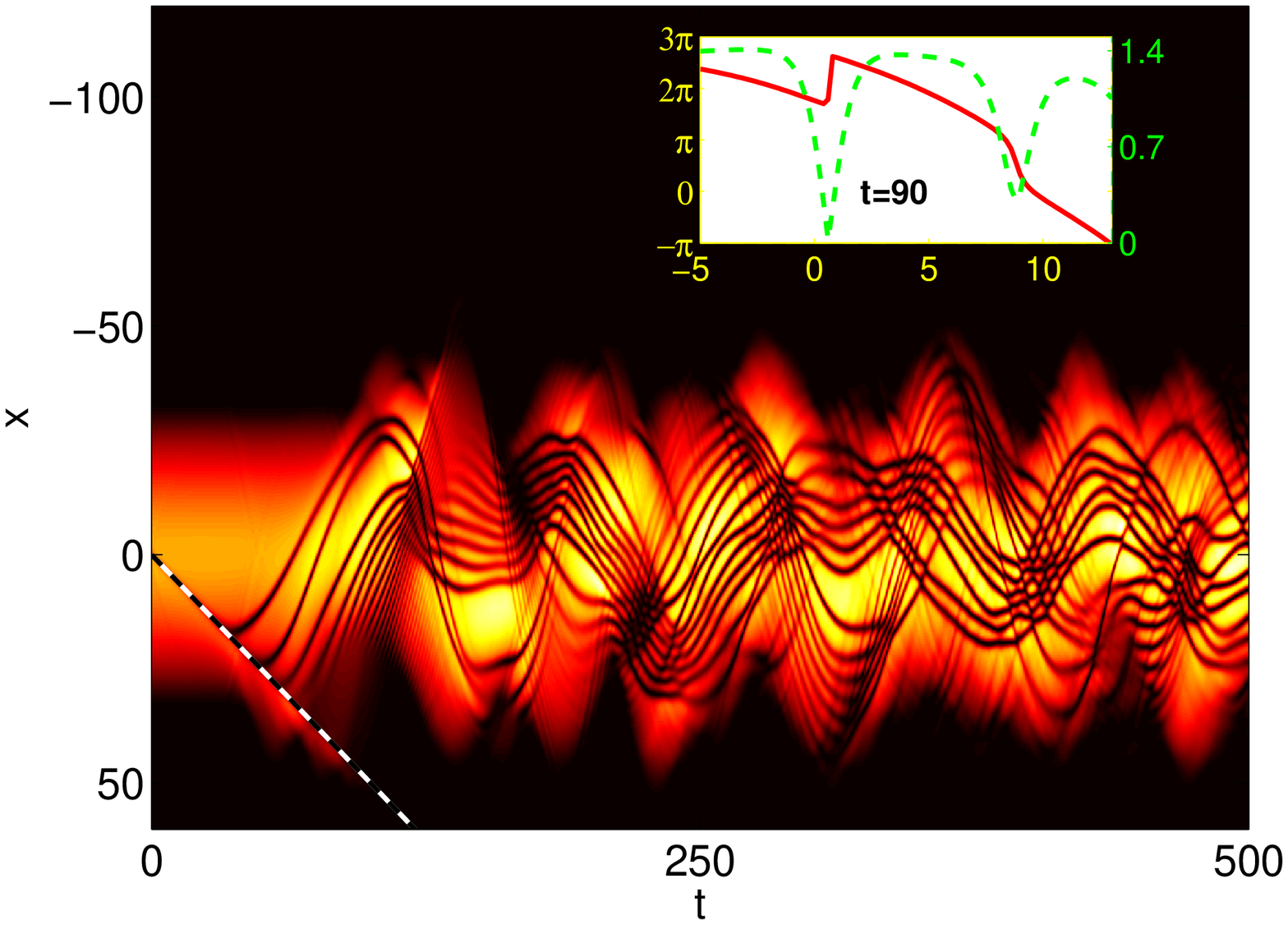}
\caption{(Color online)
Nucleation of dark solitons by a dragging defect
traversing a nodeless cloud. Depicted are the
density plots of the nodeless cloud as it is
traversed by the dragging defect.
The dashed line represents the trajectory of the defect.
The case depicted in the top panel corresponds to
a tighter confinement with trap frequency $\Omega=1$,
and parameter values
$\alpha=1.0$, $\sigma=0.35$ and $x_m=2.8$
whereas $V_0=8$, $v=0.9$ and $\epsilon^2=4$.
The bottom panel corresponds to a weaker harmonic trapping
with trap frequency $\Omega=0.04$, and
$\alpha=0.1$, $\sigma=0.08$ and $x_m=44.5$, while
$V_0=1$, $v=0.5$ and $\epsilon^2=0.16$.
The insets show the profile ([green] dashed line) and its 
corresponding phase ([red] solid line) at the indicated times.
%
}
\label{fig:def1}
\end{center}
\end{figure}

\begin{figure}
\begin{center}
    \includegraphics[width=\figwidth]{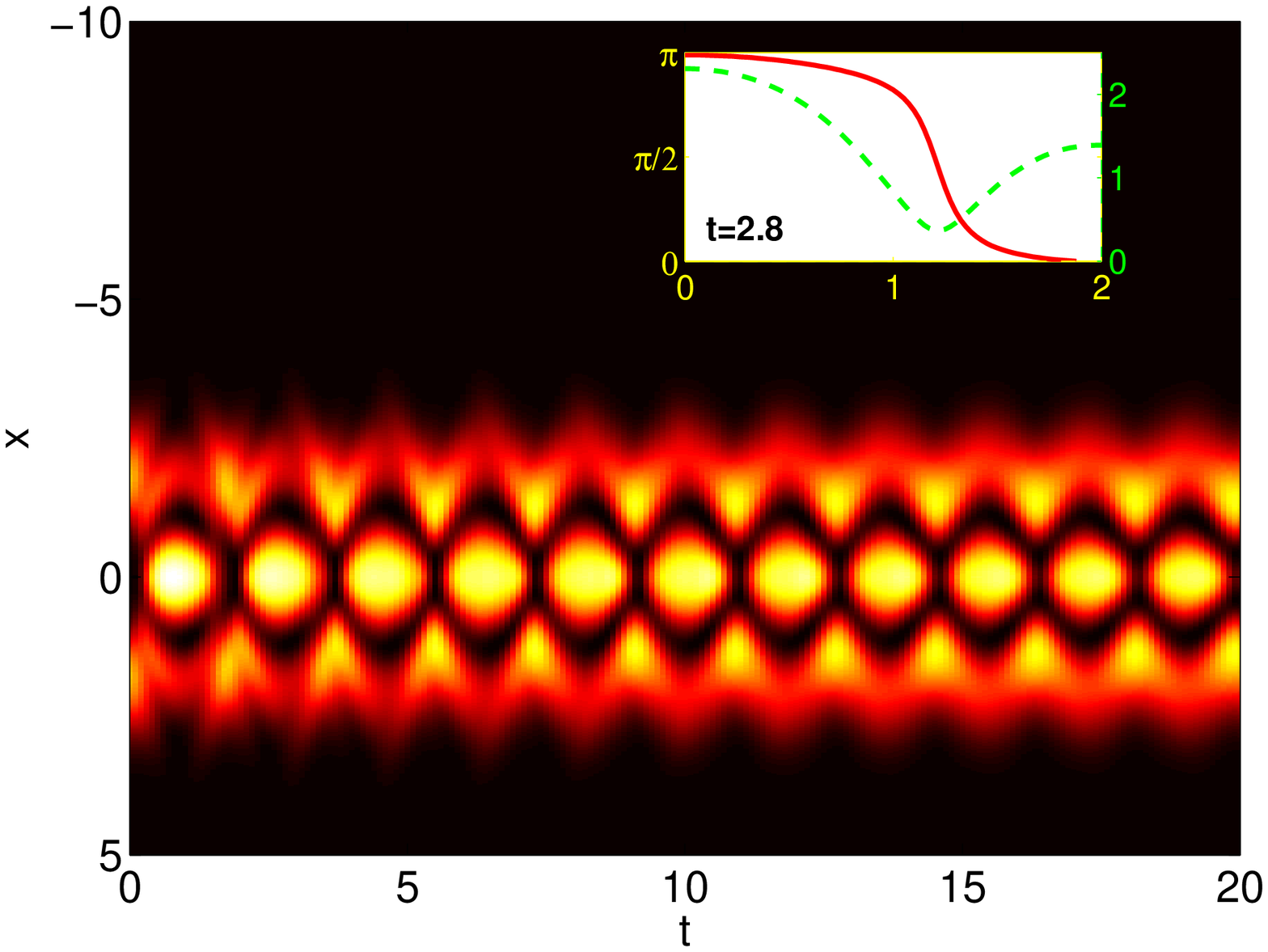}
    \includegraphics[width=\figwidth]{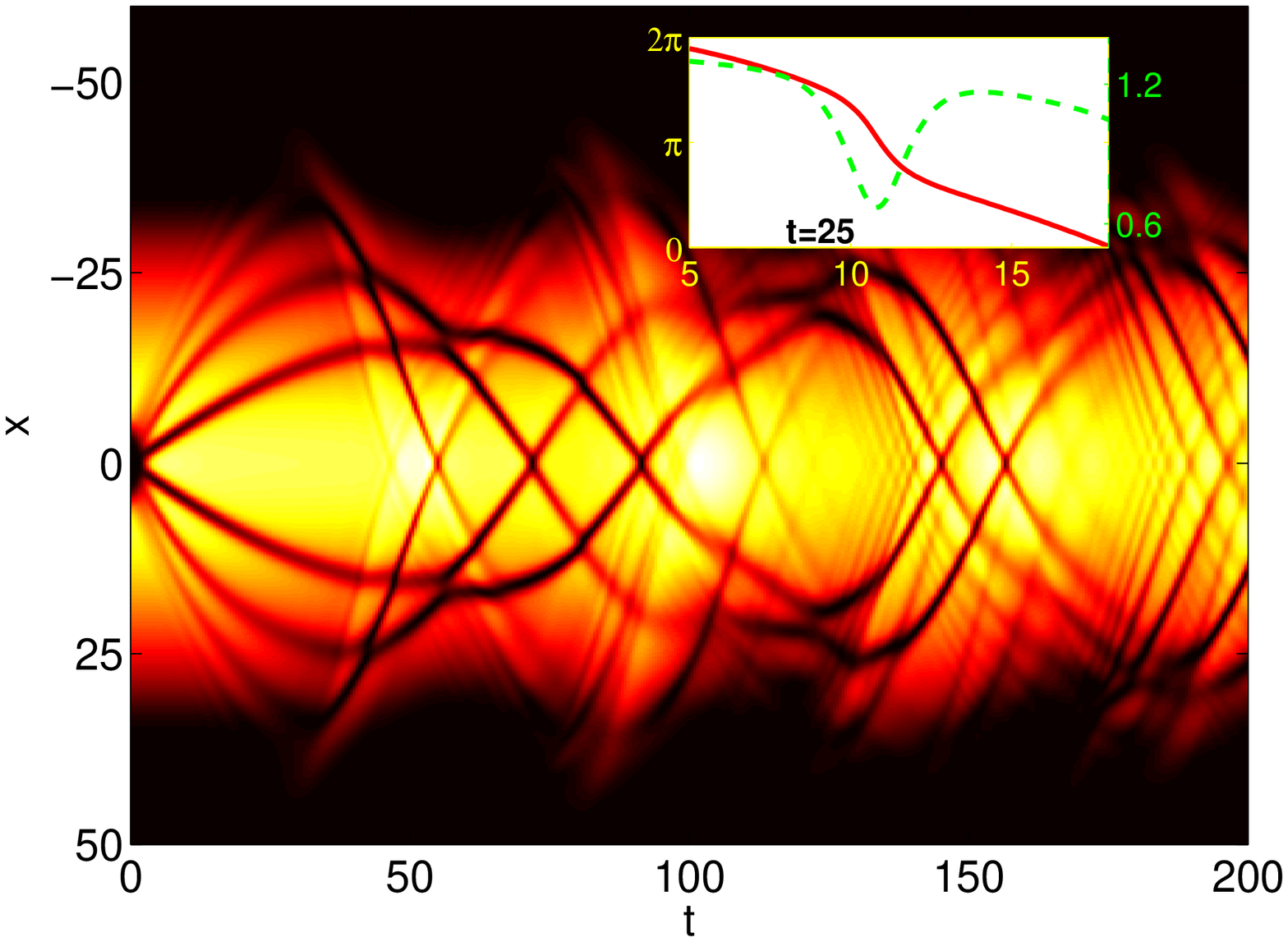}
\caption{
Nucleation of dark soliton pairs by nonlinear interference of
colliding condensate
fragments. The plots depict the density of the evolution of a
cloud created in a double well and allowed
to interfere by removing the central barrier.
Parameter values are: (top) $\alpha=1.0$, $\sigma=0.35$ and $x_m=2.8$
whereas $V_0=17$ and $\epsilon^2=1$, $\Omega=1$;
(bottom) $\alpha=0.1$, $\sigma=0.08$ and $x_m=44.5$
whereas $V_0=4$ and $\epsilon^2=16$,
while the trap frequency is $\Omega=0.04$.
The insets show the profile ([green] dashed line) and its 
corresponding phase ([red] solid line) at the indicated times.
%
}
\label{fig:dw1}
\end{center}
\end{figure}

The effects of varying the parameters $\sigma$ and $x_m$,
are depicted in Figs.~\ref{fig:sigma1} and \ref{fig:sigma2}, 
for fixed pumping spot strengths $\alpha=2$ and $\alpha=3$, respectively.
A new dynamical feature that arises for $\alpha=3$ and sufficiently large
$\sigma$ and $x_m$ (see Fig.~\ref{fig:sigma2})
is the emergence of a highly asymmetric ``sawtooth'' structure
(notice that for this value
of $\alpha$, no stable DSs are found).
Figure \ref{fig:dynfinger} shows how these
sawtooth structures emerge
from the NC when the latter is unstable. The top panel shows the
the emergence of a breathing sawtooth structure from an unstable NC. 
Notice that the asymmetry of this configuration is produced by
the amplification of the asymmetric (small) perturbation added
to the NC in order to manifest its dynamical instability.
We have checked that this breathing behavior persists
for very long times (larger than several thousands)
without any apparent decay. This is due to the fact
that, for these parameter values, the steady state sawtooth
configuration is unstable (see the spectrum depicted in the inset
of the top panel in Fig.~\ref{fig:dynfinger}) and thus the
cloud cannot decay to this state.
For other parameter values, the emerging
sawtooth structures settle to stationary
configurations without any breathing as can be seen seen
in the bottom panel of Fig.~\ref{fig:dynfinger}.
This case corresponds to parameter values inside
the elongated island depicted in the bottom panel of 
Fig.~\ref{fig:sigma2} where these sawtooth configurations
are {\em stable}.

The top panel of Fig.~\ref{fig:sigma2} depicts examples of 
stationary sawtooth structures.
Similar to the saturation of
the nodeless cloud size for large pumping spot size $x_m$,
we have also observed a saturation of the size for the
sawtooth structures for large $x_m$ (results not shown here).
It is extremely interesting that the polariton condensate
is able to support stable {\em asymmetric} sawtooth-patterned states.
Such solutions are not possible in atomic BECs.
Furthermore, it is important to mention that, because of
the left-to-right symmetry of our system, the sawtooth
states appear in pairs (left- and right-handed sawtooth structures).
In this work, for simplicity, we chose to only depict
one family as the other
one is symmetric with respect to the center of the cloud.

Finally, although our results already suggest that
nonlinear excitations in the form of dark solitons
should spontaneously emerge in polariton BECs,
we offer some alternative dynamical schemes for producing such excitations
inspired by experimental realizations within their atomic BEC
counterparts \cite{engels,weller}, which also appear to be within
reach for the case of polaritons; see e.g.~the very recent work
of Ref.~\cite{aamo} and references therein.
One possible nucleating mechanism for dark solitons is
by {\it dragging an obstacle} ---in the form of the potential of Eq.~(\ref{obs})--- 
sufficiently fast
through the condensate (see Refs.~\cite{engels} and \cite{aamo}
for relevant experimental observations in atomic and polariton condensates,
respectively).
Examples of this effect are shown in Fig.~\ref{fig:def1}.
For the relatively strong harmonic confinement considered in Eq.~(\ref{eq:dyn}),
we have found that
at most two DSs can survive, due to their relatively large size.
In this case, as shown in the top panel
of Fig.~\ref{fig:def1}, although three (or even four) DS can be seen being emitted
out of the defect path, through collisions they eventually decay down to two, that
continue to interact. The DS nature of these structures can be seen through the
phase jump of $\pi$ shown  (together with its profile) in the inset.
%
Chains of dark solitons
(alias ``dark soliton trains'') can
be produced by this dragging defect mechanism if one chooses a weaker
harmonic trap $\Omega^2 x^2/2$ [instead of
$x^2$ as in Eq.~(\ref{eq:dyn})], with a trap strength $\Omega$ sufficiently small.
If fact, as depicted in the bottom panel of Fig.~\ref{fig:def1},
a weaker trapping with $\Omega=0.04$
(corresponding to a considerably wider condensate) allows for the
formation of a train of DSs that propagates initially in the opposite direction
of the dragging defect.
However, contrary to the Hamiltonian
case of atomic BECs where the distance between the generated DSs appears to be approximately
constant
\cite{CK07}, here
DSs interact strongly with each other
and with the background cloud, setting it in oscillation. This might also be due to the
unstable nature of the underlying NC cloud for this parameter set. Again the DS nature
is visible from the $\pi$ phase jump at each of the nonlinear excitations shown
in the inset.

Another possible nucleation mechanism for DSs is by nonlinear interference (see, e.g., Ref.~\cite{djf}
for a discussion in the context of atomic BECs).
This mechanism can be realized by
splitting the condensate into two
fragments
by adiabatically inducing a potential barrier in the
central portion of a stationary nodeless cloud,
and subsequently releasing the fragments by suddenly removing the
barrier that separates them.
This ``nonlinear interference method'' was used in atomic BEC experiments
to demonstrate the generation of vortex structures \cite{brian}
(see also theoretical work in Ref.~\cite{OurFragments})
and dark solitons \cite{weller,jeff,technion}.
Figure \ref{fig:dw1} depicts two samples of this nucleation mechanism
where two polariton condensates, initially placed in a double well potential,
are released in the harmonic trap and allowed to
interfere.
Again we show the scenarios of strong (top) and weak(er) harmonic confinement (bottom).
We considered an initial
situation where a barrier of the same shape as (but wider than) the
obstacle considered
before [cf.~Eq.~\ref{obs})] is superposed at the center of the
harmonic trap. At $t=0$ the barrier is removed,
thus allowing the two portions of the condensate to mix and interfere. As can be seen
in the top panel of  Fig.~\ref{fig:dw1},
a pair of dark pulses is
formed; but contrary to the case of the soliton trains nucleated by the dragging
defect, we observe that the DSs nucleated by the nonlinear interference mechanism
move relatively fast away from each other towards the edge of the condensate.
Notice that, as again shown in the inset, the phase jumps of
approximately $\pi$ indicate that these are pairs of genuine DSs.
The bottom panel shows the case of weaker confinement where in this case more than
one pair of nonlinear excitations is produced by the nonlinear interference.
%


\section{Conclusions}

In this work, we considered and studied a complex Gross-Pitaevskii model describing the
quasi-one-dimensional dynamics of
polariton condensates.
Our motivation was to understand the fundamental differences
between
polariton condensates and their atomic (Hamiltonian BECs)
counterparts.
We found that the
specific pumping and damping terms that have been argued as being
relevant to
polariton condensates case offer a wide range of unexpected
features when compared to atomic condensates.

The fundamental nodeless state
(regarded as the ``ground state'')
of the system was found to become unstable through a variety of mechanisms,
while excited states ---in the form of single- or multiple-dark solitons---
were found to result from the instability of the nodeless state.
The
fundamental excited state,
namely the single dark soliton
was also found to be subject to instabilities,
leading to either spontaneous formation of multi-dark soliton states or even
to emergence of a
``dark-soliton-turbulence'' (when highly unstable). All these are
dynamical manifestations which significantly distinguish the polaritonic
case
from the atomic BEC variant of the problem, yet we have intuitively
attributed them to the emerging competition of length scales (among
the intrinsic length scale of the trapped system and the length scale
of the applied forcing).
We also observed the emergence of {\em stable asymmetric} sawtooth-like
configurations which are not present in atomic BECs.

Finally,
other techniques, namely nucleation of dark solitons by dragging
an obstacle through the condensate and through 
nonlinear interference of colliding condensate fragments,
have been studied. These were adapted from the atomic condensate case in order
to produce fundamental nonlinear excitations ---in the form of dark solitons.
It was shown that both techniques were efficient in doing so.
%

There are numerous interesting avenues that this work opens in the
way of future directions. On the one hand, it would be
useful to try to develop analytical tools to understand the
instability of dark solitons in polariton condensates, as well
as that of the backgrounds (i.e., the nodeless clouds) on which
these ``live''. On the other hand, it would be interesting to extend
some of the present considerations, such as the spectral analysis
of stationary states, or the nonlinear interference technique for
producing coherent structures in the 2D case to examine some of
the theoretical and potentially even experimentally relevant (in the latter
case) results thereof. Such studies are currently in progress
and will be reported in future publications.

\section*{Acknowledgments}
J.C. acknowledges financial support from the MICINN project FIS2008-04848.
A.S.R.\ gratefully acknowledges the hospitality from 
the Department of Mathematics of the University of Massachusetts,
financial support from FCT through grant SFRH/BSAB/1035/2010, and the use of
computational resources from GOE-Inesc Porto.
R.C.G.\ gratefully acknowledges the hospitality
of the Grupo de F\'{\i}sica No Lineal (GFNL, University of Sevilla, Spain)
and support from NSF-DMS-0806762, Plan Propio de la Universidad de Sevilla,
Grant No. IAC09-I-4669 of Junta de Andalucia and Ministerio de Ciencia e
Innovaci\'on, Spain.
P.G.K.\ acknowledges the support from NSF-DMS-0806762 and from the
Alexander von Humboldt Foundation.
The work of D.J.F. was partially supported by the Special Account for
Research Grants of the University of Athens.

\end{document}